\documentclass[12pt]{article}
\usepackage{amssymb,amscd,amsmath,amsthm}
\usepackage{latexsym,amstext}
\usepackage{latexsym,amstext}
\usepackage{color}
\usepackage{graphicx}
\usepackage{epstopdf}
\usepackage{cite}
\begin{document}

\title{\bf A study of  periodic potentials based on quadratic splines}

\author{M. Gadella$^1$, L.P. Lara$^2$. \\  Departamento de F\'isica Te\'orica, At\'omica y Optica and IMUVA \\ $^1$Universidad de Va\-lladolid, 47011 Valladolid, Spain \\$^2$ Universidad Nacional de Rosario and IFIR, Rosario, Argentina}

\maketitle

\begin{abstract}

We discuss a method based on a segmentary approximation of solutions of the Schr\"odinger by quadratic splines, for which the coefficients are determined by a variational method that does not require the resolution of complicated algebraic equations. The idea is the application of the method to one dimensional periodic potentials. We include the determination of the eigenvalues up to a given level, and therefore an approximation to the lowest energy bands.   We apply the method to concrete examples with interest in physics and discussed the numerical errors.

\end{abstract}

\section{Introduction}

In ordinary quantum mechanics, one dimensional potentials serve as a laboratory to test properties of quantum systems. Also, one  dimensional periodic potentials serve as models to study crystals in solid state and condensed matter. It is well known that the number of exactly solvable periodic systems is small. The main objective of the present paper is to propose a method to obtain approximate eigenvalues and eigenfunctions for one dimensional models with a periodic potential along a discussion on its precision and a comparison with other available methods. This method is particularly simple to be implemented. 

We shall reach our objective in two steps. First, we use a quadratic segmentary interpolation by quadratic splines and then, a numerical integration of second order differential equations. 

There exist a number of well known algorithms for the determination of quadratic splines \cite{KIN,ABR}. In general, they need of the resolution of algebraic equations. The originality of our method lies on the introduction of a variational tool that minimizes the oscillations due to the interpolating polynomial. In addition, the coefficients of this polynomial can be explicitly determined through elementary functions. The procedure avoids the need for the resolution of those algebraic equations that  usual methods require. 

Since we are mainly interested  in the solution of one dimensional periodic systems in non-relativistic quantum mechanics, we intend to give approximate solutions to the eigenvalue problem for equations of the type

\begin{equation}\label{1}
y''(x)+(E-V(x))y(x)=0\,.
\end{equation}

We are aware of the existence in the literature a wide variety of methods to reach this objective. See for instance, \cite{BU,HAI}. We should also mention those methods based on quadratic and cubic splines \cite{NA,RAVI}. 

We apply this method to obtain approximate eigenvalues and eigenfunctions for some choice of potentials that may have interest in physics as well as some methodological and pedagogical interest. In this manuscript, we choose the potential $V(x)=|x|^s$ for $s>0$. We consider this potentials as defined on the interval $[-1,1]$ and then, copies by periodicity on intervals $[2n-1,2n+1]$ for $n=0, \pm 1,\pm2,\dots$. Observe that for $s=2$, it is exactly solvable. These potentials show relative minima at the origin and periodically spaced points. This suggests that a good approximation for $V(x)$ would be by means of quadratic splines \cite{RLZ}. As matching conditions, we use those given in standard Kronig-Penney models \cite{KP,AM}. 

It is interesting to remark that as $s\longmapsto\infty$, our periodic potential converges pointwise to the Dirac comb. This is a sequence formed by equally spaced Dirac deltas supported at the points $2n+1$, $n=0,\pm 1,\pm 2\,\dots$. This periodic potential has been studied \cite{COR,LAR,AK,KU}. 

Once we have obtained the eigenvalues of \eqref{1}, we have determined the energy bands corresponding to the given periodic potential. 

When designing a new method that obtains results that could be reach by another method, we are guided by two main motivations: simplicity and precision. We believe that our method is simpler than others in terms of algebraic operations and CPU times. The precision is something that should be measured by reliable criteria. These criteria do exist in the literature. Only we need to compare the precision of our method by comparison with other methods, if possible within the framework of some exactly solvable examples. For this comparison, we have chosen the Mathieu equation. 

We have discussed the spline method with our variational determination of the coefficients of the segmentary solution in Section 2, which includes an analysis on the control of oscillations. The determination of eigenvalues and hence of the energy bands follows the lines described in Section 3. In section 4, we control the quality of our method using the Mathieu equation as reference. Section 5 is devoted to one dimensional models of interest in quantum physics. We consider the potential $V(x)=|x|^s$, first confined on an interval and then, extended by periodicity to all values of the real line. A discussion on the precision by error estimations are given in Section 6.

\section{Quadratic splines and variational method}

The aim of the present Section is to offer the reader a comprehensible presentation of the approximation method, based on quadratic splines,  that we shall use in our study. 

Let us consider a finite interval $[a,b]$ of the real line, in which we select $n+1$ nodes, $x_0=a<x_1<x_2<\dots<x_n=b$. We call the difference between two successive nodes $h_k:=x_{k+1}-x_k$. Take a continuous function $y(x):[a,b]\longmapsto \mathbb R$ and denote the values of $y(x)$ at the nodes as $y_k:=y(x_k)$, $k=0,1,\dots,n$. A quadratic segmentary interpolator, $S(x)$,  for the function $y(x)$ is a continuous function $S(x):[a,b]\longmapsto\mathbb R$ with continuous first derivative such that:

\smallskip

1.- On each interval of the form $I_k=[x_{k-1},x_k]$, $k=1,2,\dots,n$, $S(x)\equiv P_k(x)$, where $P_k(x)$ is a polynomial of order two. 

\smallskip

2.- The function $S(x)$ interpolates $y(x)$. This means that for any of the nodes $\{x_k\}$, one has

\begin{equation}\label{2}
P_k(x_{k-1})=y_{k-1}\,,\qquad P_k(x_k)=y_k\,,\qquad k=1,2,\dots,n\,.
\end{equation}

\smallskip

In addition, we demand that the derivative $S'(x)$ be continuous on the interval $[a,b]$. This means that it must be continuous at the internal nodes, so that

\begin{equation}\label{3}
P'_k(x_k)=P'_{k+1}(x_k)\,,\qquad k=1,2,\dots,n-1\,.
\end{equation}

The expressions contained in \eqref{2} and \eqref{3} provide of $3n-1$ equations. Since each of the polynomials $P_k(x)$ is of degree two, it is determined as we fix its three coefficients. Since the interpolation requires  the knowledge of $n$ polynomials, we have $3n$ real indeterminates, which are the coefficients of $P_k(x)$ for all $k$. This means that the function $S(x)$ is not determined yet. We still need another condition to be fixed later.

In order to simplify our expressions, we shall assume in the sequel that the distance between nodes remains constant and solely depends  on the order of partition, so that $h\equiv h_k=\frac{b-a}{n}$, $k=1,2,\dots,n$. This condition is not essential as it does not produce any variation in our results.

There are usual methods to determine these $3n$ coefficients, which rely on the resolution of linear algebraic systems \cite{KIN,PRESS}. {\it We here introduce a method that precisely avoids the resolution of algebraic systems}. It goes as follows: on each interval $[x_{k-1},x_k]$, we consider the Lagrange polynomial $p_k(x)$, defined as:

\begin{equation}\label{4}
p_k(x)=\frac{x-x_{k-1}}{h}\,y_k-\frac{x-x_k}{h}\,y_{k-1}\,.
\end{equation} 

Obviously, these Lagrange polynomials satisfy the relations $p_k(x_{k-1})=y_{k-1}$, $p_k(x_k)=y_k$, $k=1,2,\dots,n$. Then, we define the polynomial $P_k(x)$ on this interval as

\begin{equation}\label{5}
P_k(x):= p_k(x)+a_k (x-x_{k-1})(x-x_k)\,, \qquad k=1,2,\dots,n\,.
\end{equation}

Then, use \eqref{5} in \eqref{3}. Simple algebraic manipulations give:

\begin{equation}\label{6}
a_{k+1}=\frac{y_{k+1}-2y_k+y_{k-1}}{h^2}-a_k\,,\qquad k=1,2,\dots,n-1\,.
\end{equation}

Obviously, equation \eqref{6} determines each of the $a_k$ in terms of $a_1$. We have mentioned that we need to give an extra condition in order to determine the interpolating function $S(x)$. This new condition is a recipe to fix $a_1$. 

We obtain the relation between $a_k$ and $a_1$ through some algebraic manipulations. The relation for $k=2$ comes straightforwardly from \eqref{6} and is given by

\begin{equation}\label{7}
a_2=-a_1+\frac{y_0-2y_1+y_2}{h^2}\,.
\end{equation}

In general, we have that

\begin{equation}\label{8}
a_k=(-1)^{k+1}\,a_1+r_k\,,\qquad k\ge 2\,,
\end{equation}
with

\begin{equation}\label{9}
r_1=0\,,\qquad r_2= \frac{y_0-2y_1+y_2}{h^2}\,,\qquad r_3=-\frac{y_0-3(y_1-y_2)-y_3}{h^2}\,.
\end{equation}

Finally, for $k\ge 4$, we have that

\begin{equation}\label{10}
 r_k=\frac{(-1)^k}{h^2} \left[y_0-3(y_1+(-1)^ky_{k-1})+(-1)^ky_k+ 4\sum_{j=2}^{k-2} (-1)^j y_j   \right]\,.
\end{equation}

Once we have fixed $a_1$, we have an explicit expression for the coefficients of the interpolating function $S(x)$ in terms of elementary functions. In general, this is not the result obtained through the standard methods based on the solution of an algebraic system \cite{FS,RA,PRESS}.

Then, it is now the right time to fix $a_1$. It is desirable to control the oscillations of the interpolating function $S(x)$ and this could be performed by a sort of variational method. First of all, we define an {\it segmentary linear interpolating function}, $s(x)$, which in each interval $I_k:=[x_{k-1},x_k]$ is equal to $p_k(x)$, as given by \eqref{3}. Thus, the function $s(x)$ is completely determined on the whole interval $[a,b]$, while $S(x)$ is given by polynomials whose coefficients depend on $a_1$, which is undetermined so far. 

Now, we determine $a_1$ so that it be a critical point (usually a minimum) of the integral

\begin{equation}\label{11}
E(a_1):= \int_{x_0}^{x_n} (S(x)-s(x))^2\,dx\,.
\end{equation}    

Recall that $a=x_0$ and $b=x_n$ are the end points of the interval under our consideration. Then, we take into account that the interpolating functions $S(x)$ and $s(x)$ are defined on each subinterval $I_k$ as \eqref{4} and \eqref{3}, respectively, to write \eqref{11} as

\begin{equation}\label{12}
E(a_1)=\sum_{k=1}^n a_k^2 \int_{x_{k-1}}^{x_n} (x-x_k)^2(x-x_{k-1})^2\,dx\,.
\end{equation} 

The integrals in \eqref{12} are trivial, so that we readily obtain that \eqref{12} is equal to

\begin{equation}\label{13}
E(a_1)=\frac{h^5}{30} \sum_{k=1}^n a_k^2\,.
\end{equation}

Since the desired value of $a_1$ is a critical value of \eqref{13}, it must be a solution of the equation given by

\begin{equation}\label{14}
\frac{d}{da_1}\,E(a_1)=0\,.
\end{equation}

Using \eqref{7} and \eqref{8}, we have that

\begin{equation}\label{15}
\frac{d}{da_1}\, a_k= (-1)^{k+1}\,,\qquad k \ge 1\,.
\end{equation}

Then, take the derivative with respect to $a_1$ in \eqref{13} and use \eqref{8} and \eqref{7}.  Due to \eqref{14}, this result has to be zero for the desired value of $a_1$, which is

\begin{equation}\label{16}
a_1=\frac 1n \sum_{k=1}^n (-1)^k\,r_k\,.
\end{equation}

Explicit expressions for $r_k$ have been given in \eqref{8}, so that if we use them in \eqref{16},  it results

\begin{equation}\label{17}
a_1=\frac 1{nh^2} \left[ (n-1)y_0-(3n-4)y_1+\sum_{j=2}^{n-1} 4(-1)^j\,(n-j)y_j+(-1)^n y_n \right]\,.
\end{equation}

Once we have determined $a_1$, we may obtain all coefficients $a_k$ using \eqref{8} and \eqref{17} under the following form:

\begin{equation}\label{18}
a_k=\sum_{j=0}^n c_{k,j}\,y_j\,.
\end{equation}

Note that, while $j=0,1,2,\dots,$, $k=1,2,\dots,n$, then, the matrix $C$ with entries $\{c_{ij}\}$ has $n$ rows and $n+1$ columns.  Let us call $Y\equiv \{y_j\}$ the column matrix whose components are the values of $y_j$, $j=0,1,\dots,n$ and $X(x)$ to the diagonal matrix with entries $(x-x_{k-1})(x-x_k)$:

\begin{equation}\label{19}
X(x)=\left( \begin{array}{cccc} (x-x_0)(x-x_1)  & 0 & \dots & 0 \\[2ex] 0 & (x-x_1)(x-x_2) & \dots & 0 \\[2ex] 0 & 0 & \dots& 0 \\[2ex] 0 & 0 & \dots & (x-x_{n-1})(x-x_n) \end{array} \right)\,.
\end{equation}

Finally, $P(x)$ and $p(x)$ are the column matrices with coefficients given by $P_k(x)$ and $p_k(x)$, as defined in \eqref{4} and \eqref{4}, respectively. Then, using the above results we obtain the following expression for the segmentary function $S(x)$, which in matrix form is

\begin{equation}\label{20}
P(x)=p(x)+X(x)\,C\,Y\,.
\end{equation}

It is relevant to underline here that $C$ depends solely on $n$ and not on the choice of the nodes $x_k$, which we  have chosen as non-equidistant, or the function $y(x)$. For instance, for $n=5$ this matrix is given by

\begin{equation}\label{21}
C=\frac 15 \left( \begin{array}{cccccc} 4 & -11 & 12 & -8 & 4 & -1 \\[2ex] 1 & 1 & -7 & 8 & -4 & 1 \\[2ex]  -1 & 4 & -3 & -3 & 4 &  -1 \\[2ex]  1 & -4 & 8 & -7 & 1 & 1 \\[2ex] -1 & 4 & -8 & 12 & -11 & 4   \end{array}  \right)\,,
\end{equation}
which is a $5\times 6$ matrix. In addition, the entries of the matrix $C$ have the following property of symmetry:

\begin{equation}\label{22}
c_{1+k,1+j}=c_{n-k,n+1-j}\,,\qquad k=1,2,\dots,m\,,
\end{equation}
$m$ being the biggest integer smaller than $n/2$. 

The result given in equation \eqref{20} provides a complete determination of the interpolating function $S(x)$. Note that we have a complete and exact expression for all coefficients given by a recurrence law. This is rather uncommon to be find by another methods and gives added value to the present contribution.   

\subsection{Oscillations control}

Using test functions, we have performed a big number of numerical experiments, which suggest that our method greatly limits oscillations, whose presence produces serious control problems in interpolation methods. Since the objective of the present paper is the study of periodic potentials based in functions of type $f(x)=|x|^\alpha$, let us use the simplest possibility, i.e., that with $\alpha=1$ to illustrate our numerical experiments. 

Then, let us take $y(x)=|x|$, on $-1\le x\le 1$ and use the interpolation polynomials  \eqref{5} with equidistant nodes, i.e.,  $h_n=2/n=h$. In this case, the interpolating function $S(x)$, constructed using any kind of Lagrange polynomials $L_n(x)$, {\it does not} converge pointwise to $y(x)$, i.e., $\lim_{n\to\infty} L_n(x)\ne y(x)$ on each interval. This is the so called the Runge paradox \cite{KIN,HEN}, which shows that the approximating polynomials for $S(x)$ cannot be constructed in general, using equidistant nodes.  

In order to evaluate the precision and quality of the interpolating function $S(x)$, we define the mean square average $e$ as

\begin{equation}\label{23}
e:= \frac 1{x_n-x_0} \int_{x_0}^{x_n} (Q(x)-y(x))^2\,dx\,,
\end{equation}
where $Q(x)$ is the approximating function we test. 

The Lagrange interpolation shows strong oscillations for $n>15$. As a sample, let us show the error for a few values of $n$:

\bigskip
$$
\begin{array}[c]{ccccc}
n & \,& 10 & 15 & 20 \\[2ex]
e & \,& 4.1\, 10^{-2} & 1.4\,10^{-2} & 3.6\, 10^2
\end{array}
$$

\bigskip

Let us compare this estimation with the estimation of errors when the interpolation uses our method. We observe that the oscillations are negligible even for rather high values of $n$:

\bigskip
$$
\begin{array}[c]{cccccc}
n &\, & 10 & 20 & 50 & 100\\[2ex]
e & \, & 1.3\,10^{-3} & 3.3\, 10^{-4} & 5.3\,10^{-5} & 1.3\, 10^{-5}
\end{array}
$$
\bigskip

Next, we use the function $f(x)=\sin 2\pi x$ defined on the interval $-1\le x\le 1$. Then, we apply three different interpolations: first a quadratic spline, then a cubic spline and, finally, the interpolation above defined. The respective errors are the following:

\bigskip
$$
\begin{array}[c]{cccccc}
n & \, & \, & 10 & 50 & 100 \\ [2ex]
e\,\text{ Quad Spline} &\, &\, & 5.0\, 10^{-4} & 1.5\, 10^{-8} & 1.5\, 10^{-10} \\[2ex]
e\, \text{ Cubic Spline} & \, & \, & 5.0\, 10^{-4} & 4.1\, 10^{-11} & 9.0\, 10^{-14} \\[2ex]
e\, \text{ Our method} &\, &\, & 2.0\,10^{-4} & 4.5\, 10^{-9} & 5.0\, 10^{-11}
\end{array}
$$

\bigskip

In the latter table the data for quadratic and cubic splines have been obtained by means of the sofware Mathematica.

Summarizing, we have discussed a  method for quadratic interpolation, in which the coefficients for the interpolating function are determined in a very simple manner {\it without the need of solving algebraic systems}. This is one of the greatest advantages of our method. There is a second advantage, this method seems to control the error efficiently.  However, we have not obtained a precise formula to obtain the error, which is not immediate. Therefore, we have estimations of the error for given examples only. 

In the next section, we shall use these results to deal with Sturm-Liouville problems. 

\section{Eigenvalue determination.}

Along the present section, we intend to propose a method to solve the linear Sturm-Liouville problem. There exists various methods that fulfil this objective; nevertheless, we propose a simple alternative based on an algebraic  resolution. We go back to \eqref{1}, where the potential $V(x)$ is continuous on a finite interval $(a,b)$. We extend $V(x)$ to all real values of $x$ by periodicity outside the interval $(a,b)$. Then, according to the Floquet theorem, each solution should have the form:

\begin{equation}\label{24}
y(x)=e^{\lambda x}\,z(x)\,,
\end{equation}
where $z(x)$ is periodic with period $T=b-a$. Here, $\lambda$ is the Floquet exponent. According to the Floquet theorem, solutions to \eqref{1} should satisfy the following boundary conditions on the interval $(a,b)$:

\begin{equation}\label{25}
y(a)=e^{\lambda(a-b)}\,y(b)\,,\qquad y'(a)=e^{\lambda(a-b)}\,y'(b)\,.
\end{equation}

There are two Floquet exponents, $\lambda_1$ and $\lambda_2$, which are related by the equation $\lambda_1=-\lambda_2$ and are defined modulo  additive constants of the form $\frac{2n\pi }T\,i$, $n=0,\pm1,\pm 2,\dots$. For any possible eigenvalue $E$ of \eqref{1}, there exists at most two linearly independent periodic solutions. If $\lambda= 0$ the solution is periodic, if the real part of $\lambda$, Re $\lambda$, vanishes, Re $\lambda=0$, both solutions are quasi-periodic, i.e., they may be written in terms of a linear combination of harmonics such that the ratio between their frequencies is not an integer. Finally, if Re $\lambda\ne 0$ there are not bounded solutions, hence periodic.  

As is well known, when $V(x)$ is a quadratic polynomial, \eqref{1} has an explicit solution in terms of parabolic cylindric functions $D_\nu$ \cite{ABR,CHI}. Then, we propose an approximation of the potential $V(x)$, whatever it may be, by means of the quadratic splines $S(x)$ as defined in Section 2. If this is the case, we shall explicitly determine the solution using the partition that we have already used for the determination of $S(x)$. 

It seems natural to integrate \eqref{1} on the basic interval $[a,b]$, since the period is $T=b-a$. On this interval, we give a partition, such as $a=x_0<x_1<x_2<\dots,x_n=b$.  On each interval $I_k=[x_{k-1},x_k]$, $k=1,2,\dots,n$, we approximate $V(x)$ by $P_k(x)$ as defined in \eqref{5}. Consequently, for each interval $I_k$, \eqref{1} takes the following approximation:

\begin{equation}\label{26}
y''_k(x)+(E-P_k(x))y_k(x)=0\,.
\end{equation}

Since $P_k(x)=a_kx^2+b_kx+c_k$ is a quadratic polynomial on the considered interval, the solution of \eqref{26} is given as a linear combination of two independent parabolic-cylindric functions, so than on each of the $I_k$, we have that

\begin{equation}\label{27}
y_k(x)=c_{1k}D_{\nu_{1,k}}(z)+c_{2k}D_{\nu_{2,k}}(iz)\,.
\end{equation}

Here,

\begin{eqnarray}
\nu_{1,k}=\frac{-4a_k^{3/2}+b_k^2+4a_k(E-c_k)}{8a_k^{3/2}}\,,\label{28}\\[2ex]
\nu_{2,k}=-\frac{4a_k^{3/2}+b_k^2+4a_k(E-c_k)}{8a_k^{3/2}}\,,\label{29}\\[2ex]
z_x=\frac{b_k}{\sqrt 2\, a_k^{3/4}} + \sqrt 2\,a_k^{1/4}\,x\label{30}\,.
\end{eqnarray}

Let us assume that we have fixed the initial values $y_1(0)$ and $y'_1(x_0)$. This allows to determine the solution $y_1(x)$ on the initial interval $I_1$. This provides the initial values $y_1(x_1)$ and $y'_1(x_1)$, which give the solution for \eqref{26} on the interval $I_2$ and so on. Then at each node, we have some continuity conditions given by

\begin{equation}\label{31}
y_k(x_{k-1})=y_{k-1}(x_{k-1})\,,\qquad  y'_k(x_{k-1})=y'_{k-1}(x_{k-1})\,, \quad k=2,\dots,n\,.
\end{equation}

This is sufficient in order to guarantee the continuity at any order. Furthermore, relations \eqref{31} serve to determine the coefficients $c_{1k}$ and $c_{2k}$ in \eqref{27}.  In fact, since $y_k(x)$ is a linear function on its integration constants, equations \eqref{27} provide $c_{1k}$ and $c_{2k}$ as linear functions of $c_{1,k-1}$ and $c_{2,k-2}$ and hence of $c_{11}$ and $c_{21}$. Thus, $y_k$ depends on $E$, $c_{11}$, $c_{21}$ and $x$, $y_k(x)=y_k(E,c_{11},c_{21},x)$. 

We still have to determine $E$, $c_{11}$ and $c_{21}$. The coefficient $c_{11}$ may be arbitrarily chosen, so that let us write $c_{11}=1$. For $c_{21}$ and $E$, we use the matching conditions \eqref{25}.  This gives an algebraic system, which is linear on $c_{21}$ and non-linear on $E$. This is

\begin{eqnarray}
D_{\nu_{11}}(z_a)+c_{21}D_{\nu_{21}}(iz_a)=\exp(\lambda(a-b))(c_{1n}D_{\nu_{1n}}(z_b)+c_{2n} D_{\nu_{2n}}(iz_b))\,, \label{32}\\[2ex]
D'_{\nu_{11}}(z_a)+c_{21}D'_{\nu_{21}}(iz_a)=\exp(\lambda(a-b))(c_{1n}D'_{\nu_{1n}}(z_b)+c_{2n} D'_{\nu_{2n}}(iz_b))\,. \label{33}
\end{eqnarray}

From the above analysis, we conclude that for $k=2,3,\dots,n$, the coefficients $c_{ik}$, $i=1,2$, are functions on $c_{11}$ and $c_{21}$, i.e., $c_{ik}=c_{ik}(c_{11},c_{21})$. This dependence is linear. Note that, due to the arbitrariness of $c_{11}$, we have made the choice $c_{11}\equiv 1$. 

Next, we proceed as follows: From \eqref{32}, we obtain an explicit expression for $c_{21}$ in terms of $c_{1n}$ and $c_{2n}$. Then, we use this result in \eqref{33} and obtain $c_{21}$ in terms of $E$ and $\lambda$, $c_{21}=c_{21}(E,\lambda)$. Let us write this identity as

\begin{equation}\label{34}
F_\lambda (E)=0\,.
\end{equation}

This is a transcendental equation, whose solutions are the desired values of $E$. We have to solve it using numerical methods. 

Note that $F_\lambda(E)$ depends on the parabolic-cylindric functions. These functions do not have a simple form in terms of elementary functions. As a consequence, the numerical resolution of \eqref{34} may be slow in terms of CPU times. A possible solution for this problem, which unfortunately has the undesirable consequence of lost of precision,  is the span of the solution of \eqref{26} in Taylor series. Then, on each interval $I_k$, we consider the truncated Taylor series of order $m$ given by:

\begin{equation}\label{35}
y_{k,s}(x) \approx \sum_{j=0}^m \alpha_{s,j}^k (x-x_{k-1})^j\,.
\end{equation}

Here, the index $s$ stands for two linearly independent solutions, so that $s=1,2$.  We determine these two solution by means of the initial conditions $y_{k,1}(x_{k-1})=0$ and $y_{k,1}(x_{k-1})=1$ for $s=1$ and $y_{k,2}(x_{k-1})=1$ and $y_{k,2}(x_{k-1})=0$ for $s=2$. 

Once we have determined approximations \eqref{35} for the parabolic cylindric functions, we replace  $D_{\nu_{s,k}}$, $s=1,2$ by \eqref{35} into \eqref{27}, so as to obtain

\begin{equation}\label{36}
y_k(x)\approx c_{ik}\,y_{k,1}(x)+c_{2k}\, y_{k,2}(x)\,.
\end{equation}

The integration constants $c_{sk}$, $k=1,2$, are easily determined after the construction of the functions $y_{s,k}(x)$ in \eqref{36}. This gives

\begin{equation}\label{37}
c_{1k}= y'_k(x_{k-1})\,, \qquad c_{2k}=y_k(x_{k-1})\,.
\end{equation}

Once we have determined $c_{sk}$, we may obtain the values for $y_k(x)$ and $y'_k(x)$ at the point $x_k$. The result is:

\begin{eqnarray}\label{38}
y_k(x_k) \approx y'_k(x_{k-1})\,y_{k,1}(x_k)+y_k(x_{k-1})\, y_{k,2}(x_k)\,, \nonumber\\[2ex]
y'_k(x_k) \approx y'_k(x_{k-1})\,y'_{k,1}(x_k) + y_k(x_{k-1})\,y'_{k,2}(x_k)\,,
\end{eqnarray}
expression valid for $k=1,2,\dots,n$. This determines $y_k(x_k)$ and $y'_k(x_k)$ in terms of  $y_k(x_{k-1})$ and $y'_k(x_{k-1})$, which are given as linear functions of $y_0(x_0)$ and $y'_0(x_0)$.  Once we have fixed these initial conditions, we have determined our approximate solutions. We propose $y_0(x_0)=1$ and $y'_0(x_0)=u$, where $u$ is not arbitrary but determined by the boundary conditions \eqref{25} as  

\begin{equation}\label{39}
1=\exp(\lambda(a-b))\,y_n(x_n)\,,\qquad u=\exp(\lambda(a-b))\,y'_n(x_n)\,.
\end{equation}

Once we have eliminate $u$ on relations \eqref{39}, it remains an equation of the type \eqref{34} to determine the approximate values of $E$. At this point, it may be interesting to underline that the Taylor coefficients $\alpha_{s,j}^k$ in \eqref{35}  explicitly depend on $E$, so that the resulting $F_\lambda(E)$ is a polynomial on $E$. This is a nice advantage when our purpose is the numerical determination of the roots.  

Once we have introduced the method, we shall use it in the applications. The former is standard and the other are more in the spirit of our work.

\section{Mathieu equation}

In order to check the accuracy of the method, we compare our results with the results obtained for a widely studied equation of the type \eqref{1}, for which the eigenvalues are well known: the Mathieu equation. This equation is written as

\begin{equation}\label{40}
y''(x)+(r-2q\,\cos 2x)y(x)=0\,.
\end{equation}

Choose $q<1$. The expressions for the eigenvalues $r_{k,\text{parity}}$ corresponding to the few first periodic solutions $y(x)$ with period $2\pi$ are given by

\begin{eqnarray}\label{41}
r_{1,\rm{even}}(q)= 1+q-\frac{q^2}{8}-\frac{q^3}{64}-\frac{q^4}{1536}+\dots\,,  \nonumber\\[2ex]
r_{1,\text{odd}}(-q)=1-q-\frac{q^2}{8}+\frac{q^3}{64}-\frac{q^4}{1536}+\dots\,, \nonumber\\[2ex]
r_{2,\rm{even}}(q)= 4+\frac{5}{12}\,q^2-\frac{763}{13824}\,q^4+\frac{1002401}{79626240}\,q^6+\dots\,, \nonumber\\[2ex]
r_{2,\text{odd}}(q)= 4-\frac{1}{12}\,q^2+\frac{5}{13824}\,q^4-\frac{289}{79626240}\,q^6+\dots\,,\nonumber\\[2ex]
r_{3,\rm{even}}(q)= 9+\frac1{16}\,q^2+\frac 1{64}\,q^3+\frac{13}{20480}\,q^4-\frac 5{16384}\,q^5+\dots\,, \nonumber\\[2ex] 
r_{3,\text{odd}}(-q)= 9+\frac1{16}\,q^2-\frac 1{64}\,q^3+\frac{13}{20480}\,q^4+\frac 5{16384}\,q^5+\dots\,.
\end{eqnarray} 

In order to implement the comparison, we have to choose a value of $q$, let us take $q=0.2$. We obtain the following table, where the index $k$ stands for the three former characteristic values of the Mathieu equation:

\bigskip

\[
\begin{array}[c]{cccc}
k & 1 & 2 & 3\\[2ex]
r_{k,\text{even}} & 1.19487 & 4.01658 & 9.00263\\[2ex]
r_{k,\text{odd}} & 0.795124 & 3.99667 & 9.00238
\end{array}
\]

\bigskip

We have obtained the approximate values of the eigenvalues using our method, by taking three orders of approximation of the parabolic cylindric functions $D_\nu$. These orders  are $m=3,5,7$. As the segment $[a,b]$, we have chosen $[0,2\pi]$ and the values $n=50,100$. In order to evaluate the precision, we have used the relative percent error given by

\begin{equation}\label{42}
{\rm error}\,\% =\left| \frac{r_{\rm exact}-r_{\rm approx}}{r_{\rm exact}} \right|\,100\%\,.
\end{equation}

Here, $r_{\rm exact}$ is the eigenvalue obtained through \eqref{41} and $r_{\rm approx}$ is the value obtained using the method with $\lambda=0$. The results obtained for the error given in \eqref{42} are shown in the following tables:

\bigskip

\[
\begin{array}
[c]{cccc}
m=3 & 1 & 2 & 3\\[2ex]
50\, \text{ even} & 2.3\;10^{-1} & 1.1 & 2.1\\[2ex]
100\, \text{ even} & 5.8\;10^{-1} & 2.8\;10^{-1} & 5.9\;
10^{-1}\\[2ex]
50\,\text{ odd} & 3.4\;10^{-1} & 1.0 & 2.2\\[2ex]
100\,\text{ odd} & 8.6\;10^{-2} & 2.6\;10^{-1} & 5.8\;10^{-1}
\end{array}
\]

\bigskip

\[  
\begin{array}
[c]{cccc}
m=5 & 1 & 2 & 3\\[2ex]
50\, \text{ even} & 0.0 & 1.0\;10^{-2} & 1.4\;10^{-2}\\[2ex]
100\,\text{ even} & 0.0 & 1.9\;10^{-3} & 9.9\;10^{-4}\\[2ex]
50\, \text{ odd} & 7.5\;10^{-4} & 3.2\;10^{-3} & 1.7\;10^{-2}\\[2ex]
100\,\text{ odd} & 5.0\;10^{-4} & 2.5\;10^{-4} & 1.0\;10^{-3}
\end{array}
\]

\bigskip

\[
\begin{array}
[c]{cccc}
m=7 & 1 & 2 & 3\\[2ex]
50\, \text{ even} & 8.0\;10^{-4} & 0.0 & 1.0\;10^{-4}\\[2ex]
100\, \text{ even} & 0.0 & 0.0 & 0.0\\[2ex]
50\,\text{ odd} & 5.0\;10^{-4} & 0.0 & 1.1\;10^{-4}\\[2ex]
100\,\text{ odd} & 0.0 & 0.0 & 0.0
\end{array}\,.
\]

From the above results, we see that the error \eqref{42} is small, even for the case $m=3$ and $n=50$. We conclude that the accuracy of our method is quite satisfactory.  

Next, we want to determine the eigenfunction corresponding to the eigenvalue $r_{1,\text{even}}=1.194874$ and $q=0.2$. In our method, we use the parameters $m=5$, $n=100$. We are going to calculate the mean quadratic error $E_r$ between our discrete approximate solution $y_k(x)$ and the solution which can be obtain using the Mathieu functions $C(r,q,x)$ and $S(r,q,x)$.   The error $E_r$ is given by

\begin{equation}\label{43}
E_r:= \sum_{k=0}^n (y_k-y_{\text{exact}}(x_k))^2\,.
\end{equation}

For this data, we have obtained $E_r=10^{-12}$. Nevertheless, this is just an example of what happens in general, so that for all other eigenvalues we obtain similar precision. Using a Fourier interpolation \cite{ABR} for $y_k$, we may determine the approximate eigenfunction with two harmonics as:

\begin{equation}\label{44}
y=1.02608\,\cos x-0.0262979\,\cos 3x\,.
\end{equation}

The resulting mean quadratic error is $E_r=10^{-6}$, a very satisfactory result. 

\section{The potential $V(x)=|x|^s$}

Now, we are going to apply the method to some one dimensional periodic models. This potentials may be of some interest in solid state \cite{KP,AM}. 

Since we are dealing with approximation and numerical methods, we need to specify the value of $s>0$ for concrete applications.  Let us start with the simple case $s=2$, so that its restriction to each subinterval of the form $[a,b]$ is a harmonic oscillator:

\begin{equation}\label{45}
V(x)=x^2\,.
\end{equation}

This is valid on the interval $-1\le x\le 1$ having the same shape for any other interval $[2n+1,2n+3]$, $n=0,1,2\dots$. The general solution for \eqref{45} has the form $y(x=c_1\,y_1(x)+c_2\,y_2(x)$, where the linearly independent solutions are given by

\begin{equation}\label{46}
y_1(x)= D_{\frac12\,(E-1)}(\sqrt 2\,x)\,, \qquad y_2(x)= D_{-\frac12\,(E-1)}(i\sqrt 2\,x)\,.
\end{equation}

Now, we follow the procedure given in Section 3: First of all, we use the boundary conditions \eqref{25} and, then, obtain an explicit expression for $F_\lambda(E)=0$ as in \eqref{34}. This transcendental equation should be solved numerically in order to determine the approximate values of $E$. 

As we have mentioned at the beginning of Section 3, for any given value of $E$, we have two values of $\lambda$, except for the case $\lambda=0$, where we have periodic solutions. In this latter case and starting with the explicit solution, we determine the values $E_k$ for the lower values of $k$. Next, we approximate $y(x)$ by means of the Taylor expansion of degree $m=5$, as described in Section 3. The partition of the interval $[-1,1]$ has $n=100$ segments of equal length. The eigenvalues obtained after formulas \eqref{25} and \eqref{46} coincide with those generated with the method introduced here at least in the first six digits:

\bigskip

\[
\begin{array}[c]{cccc}
k & 1 & 2 & 3\\[2ex]
E_{\rm even} & .324942 & 10.2601 & 39.8253\\[2ex]
E_{\rm odd} & .324942 & 10.1512 & 39.7994 \,.
\end{array}
\]

\bigskip

Let us go to imaginary values of $\lambda$, which as remarked before correspond to quasi-periodic solutions. In Figure 1, we represent the values of the energy in terms of the real number $\alpha$ with $\lambda=i\alpha$. As is well known, the energy bands correspond to unbounded solutions. In our case, the lowest energy bands corresponds to the segments $[2.59692,3.0]$, $[10.1512,10.2601]$, $[22.5177,22.5645]$, $[39.7994,39.8253]$. In Figure 2, we have a closer look to these bands. The symmetry which is appreciated in this Figure comes from the fact that the sum of both Floquet exponents is equal to zero. Since these coefficients are determined save for addition of $\frac{2n\pi}{L}\,i$, this picture is extended by periodicity to all real values. We recall that for non vanishing energies the eigenstates are doubly degenerated. 

Once we have analyzed the simplest case of a quadratic periodic potential, we may study the most general situation for which $s$ is arbitrary, so that

\begin{equation}\label{47}
V(x)=|x|^s\,,\qquad s>0\,,
\end{equation} 
where the interval $[a,b]$ is again $[-1,1]$ and we extend \eqref{44} to any interval $[2n+1,2n+3]$, $n=0,\pm 1,\pm2,\dots$ by periodicity. 

\begin{figure}
\begin{center}
\includegraphics[width=0.6\textwidth]{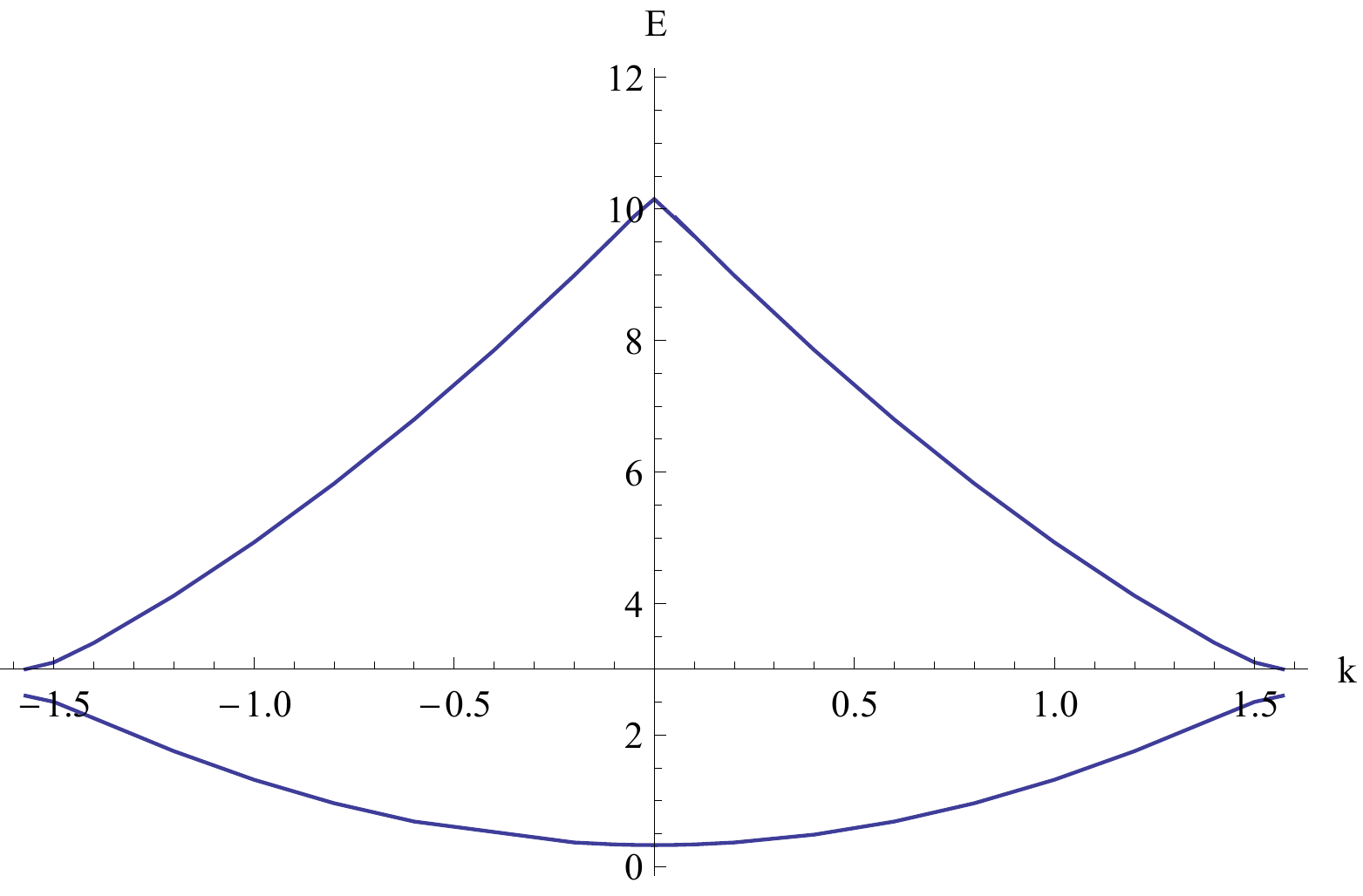}
\caption{. }
\end{center}
\end{figure}

We have obtained  a series of results for various values of $s=1/50,1/20,1/3,$ $1,2,8,20,50$. We have determined the two lower eigenvalues in terms of $\lambda$.  In Figure 3, we show our results. Observe that as $s$ increases, the curve $E^\pm_s(\lambda)$, where the signs plus and minus refers to the upper and lower curve, respectively, goes down. From this behavior, one may conjecture that for high values of $s$, $E_s(\lambda)$ goes to a limit curve $E^*(\lambda)$. On the other hand, as $s$ goes to zero we may also conjecture that $E_s(\lambda)$ tends toward another limit curve, $E^{**}(\lambda)$. In addition, the prohibited band width, $E_s^+(\pi/2)-E^-_s(\pi/2)$ decreases as $s$ grows.

\begin{figure}
\begin{center}
\includegraphics[width=0.6\textwidth]{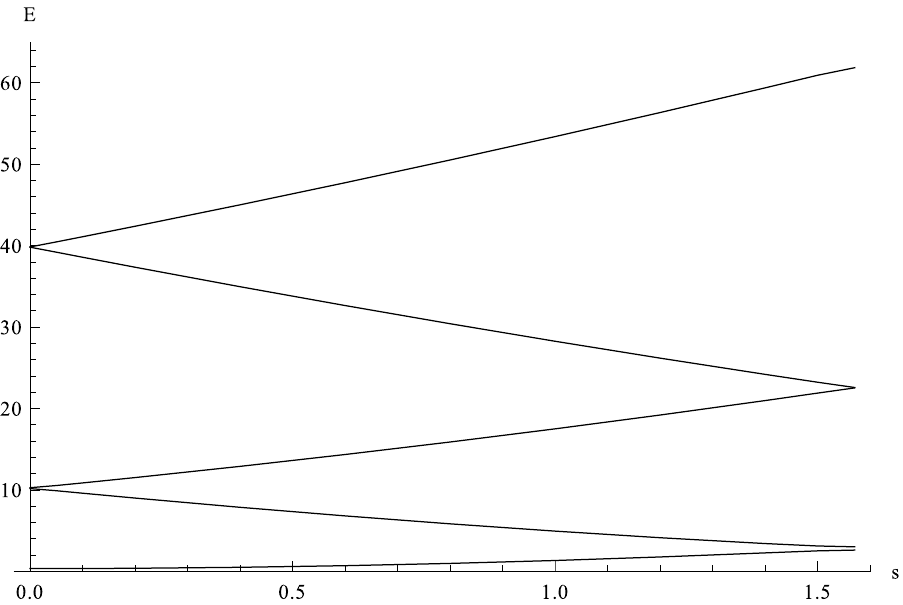}
\caption{. }
\end{center}
\end{figure}

\begin{figure}
\begin{center}
\includegraphics[width=0.6\textwidth]{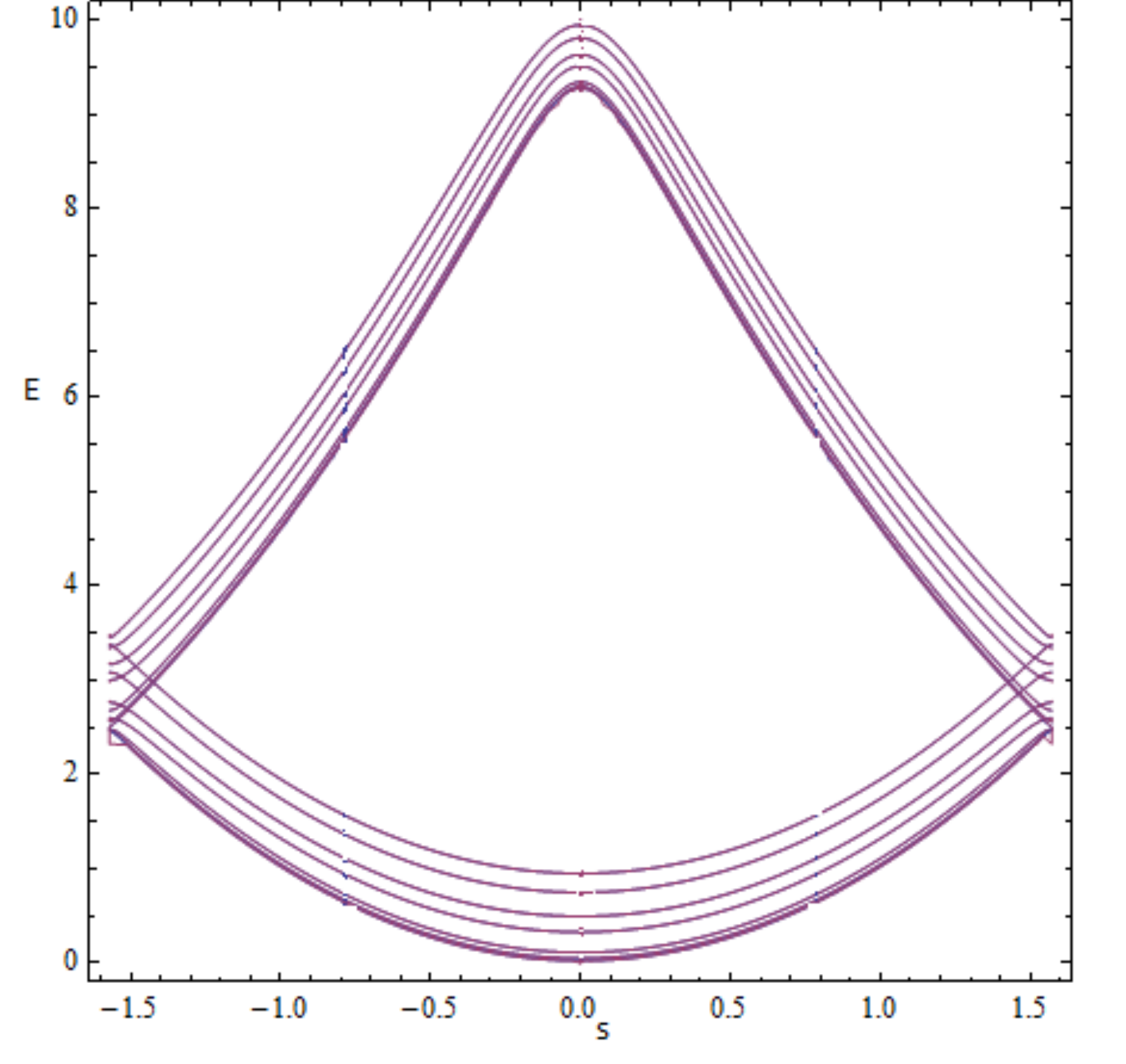}
\caption{. }
\end{center}
\end{figure}

As a particular example, let us give our results for the potential $V(x)=x^4$. Use $m=5$ and $n=100$ and let us find out the periodic solution, here $\lambda=0$, for the energy $E=10.1590085$. Let us proceed to calculate the mean quadratic error between the discrete solution, as obtained by our method, and the solution as given using the Runge-Kutta method. The result is $E_r=4.9\times 10^{-13}$. It is noteworthy to say that with other eigenvalues, we have obtain the same degree of precision. 

In addition, we have interpolating $\{y_{k}\}_{k=0}^{n}$ by means of Fourier interpolation and have obtained the following result:

\begin{eqnarray}\label{48}
y(x)= 0.0157801+0.993152\cos(\pi(1+x))-0.00683514\cos(4\pi(1+x))\nonumber\\[2ex]
-1.38926\times 10^{-4}\cos(9\pi(1+x))-3.86941\times 10^{-4}\cos (16\pi(1+x))\nonumber\\[2ex]
-1.48414\times 10^{-4}\cos(25\pi(1+x))\,,
\end{eqnarray}
the error being given by $E_r=4.6\times 10^{-3}$. Thus, \eqref{48} is the eigenfunction for $\lambda=0$ and $E=10.1590085$, which have been obtained by choosing $m=5$ and $n=100$.

\section{Error analysis}

The purpose of this section is a study of the error that emerges when using our method as described in Sections 2 and 3. The first comparison should be made using the quadratic potential $V(x)=x^2$, since this is exactly solvable. Let us consider the integration interval $[-1,1]$ and divide it in $n$ intervals of equal length.  From \eqref{27} and the form of parabolic cylindric functions, or from \eqref{36}, we determine approximate eigenvalues and their corresponding segmentary eigenfunctions. With the use of parabolic cylindric functions, we integrate \eqref{1} with periodic conditions at the points $x=\pm 1$. Henceforth, we call this solution $w_{\rm exact}(x)$, which will be used as test function in order to quantify the quality of the segmentary solutions as obtained by our method. 

In order to determine the order of the precision, we are going to use the {\it Continuous Analytic Continuation} (CAC) procedure \cite{DA}, which has been already used in one of our previous works \cite{GLN}. This procedure is a method of segmentary integration by Taylor polynomials using periodic conditions. We call $g(x)$ to the solution obtained by either by our method or by CAC. In order to establish a precision average, we define:

\begin{equation}\label{49}
{\rm error}:= \sum_{k=1}^n (w_{\rm exact}(x_k)-g(x_k))^2\,,
\end{equation}
where $x_k$ are the nodes of the partition of the interval $[-1,1]$. 

Use the method introduced in Sections 2 and 3 with $n=100$ ($n=$ number of equally spaced intervals diving $[-1,1]$) and Taylor polynomials of orders $5$ and $7$, respectively. Let us determine the errors for the three lowest eigenvalues corresponding to even functions. The results are shown in the following table:

\bigskip

\[
\begin{array}[c]{ccccc}
\, & {\rm E} & .324942 & 10.2601 & 39.8253\\[2ex]
{\rm a} & {\rm by}\,\, D_{\upsilon} & 10^{-24} & 10^{-25} & 10^{-26}\\[2ex]
{\rm b} & {\rm order}\,\, 5 & 10^{-15} & 10^{-12} & 10^{-9}\\[2ex]
{\rm c} & {\rm order}\,\,7 & 10^{-23} & 10^{-20} & 10^{-16}\\[2ex]
{\rm d} & {\rm CAC}\,\,4 & 10^{-15} & 10^{-11} & 10^{-8}\\[2ex]
{\rm e} & {\rm CAC}\,\,5 & 10^{-19} & 10^{-15} & 10^{-12}\\[2ex]
{\rm f} & {\rm CAC}\,\,6 & 10^{-23} & 10^{-19} & 10^{-15}\\[2ex]
{\rm g} & {\rm CAC}\,\,7 & 10^{-29} & 10^{-23} & 10^{-13}\,.
\end{array}
\]

\bigskip

The rows corresponding to this table mean the following. Obviously, the first row corresponds to the three values of $E$ for which we want to find the errors. In the row a, $g(x_k)$ the errors corresponds to the segmentary solution obtained after equation \eqref{27}. We see on rows b and c the solution the errors that appear when $g(x_k)$ is the segmentary solution obtained from \eqref{36}. Errors on the rows from d to g are a consequence of using CAC with Taylor segmentary polynomials of order 4 to 7, respectively. 

It is in the row a, where the errors are due to the numerical representation and not to an effect of truncation. These errors give an upper limit to the precision, since this corresponds to the solution giving the minimum error. For the other cases, we may expect to obtain the same error for segmentary polynomials of the same degree. This is, however, not true, since operations arriving to the errors are different in these cases.  This difference between arithmetic manipulations from one method to the other results in the fact that our method has a local error one oder lower than the error resulting of using CAC. 

Being true that the use of \eqref{36} introduces an contribution to the error due to truncation, it is nevertheless advantageous the use of \eqref{36} over \eqref{27}, since in the latter the CPU times grow without an improvement in the precision. In addition, the table shows that the precision obtained with order 5 (see row b) is good enough. 

The big question is now: Since the CAC procedure shows a better precision, at least in this particular example, why we need to introduce another method? The reason lies in the advantageous simplicity of our method. In fact, starting with the parabolic cylindric , we proceed to segmentary integration of \eqref{11}, where we have approximated the potential by a quadratic spline. Therefore, we do not need to calculate derivatives as is the case with CAC, where the segmentary solution is a truncated Taylor series. The coefficients in this Taylor series are obtained by means of calculation of derivatives, which is quite often tedious. It is noteworthy that when one resorts to expand the solution into power series instead of using the basis of parabolic cylindric functions, we obtain a recurrence of the type $y_{k+1}=F(x_k,y_k,z_k)$, $y_{k+1}=G(x_k,y_k,z_k)$ and $y'(x)=z(x)$. This is a particular case of CAC, in which the potential is approximated by the spline. 

Next, we consider the Mathieu equation as given in \eqref{40} and let us take the three lowest eigenvalues for $q=0.2$ and even solution. Then, in the error formula \eqref{49}, the solution $w_{\rm exact}(x)$ corresponds to the solution obtained by means of the Mathieu functions $C(r,q,x)$ and $S(r,q,x)$ defined on the interval $[0,2\pi]$. 

Then, let us go to the table below, where $E$ refers to the energy of the three lowest eigenvalues as indicated. The row a gives the error obtained when we use as $g(x_k)$ the segmentary solution obtained after \eqref{27}. All other rows give equivalent results to those given in the previous table. These results yield to conclusions, which are the same than for the results given in the previous   table.

\bigskip

\[
\begin{array}[c]{ccccc}
\, & E & 1.19487 & 4.01658 & 9.002719\\[2ex]
{\rm a} & {\rm by}\,\,D_{\upsilon} & 10^{-12} & 10^{-13} & 10^{-11}\\[2ex]
{\rm b} & {\rm order}\,\,5 & 10^{-12} & 10^{-9} & 10^{-7}\\[2ex]
{\rm c} & {\rm order}\,\,7 & 10^{-12} & 10^{-12} & 10^{-13}\\[2ex]
{\rm d} & {\rm CAC}\,\,4 & 10^{-11} & 10^{-8} & 10^{-7}\\[2ex]
{\rm e} & {\rm CAC}\,\,5 & 10^{-13} & 10^{-12} & 10^{-10}\\[2ex]
{\rm f} & {\rm CAC}\,\,6 & 10^{-13} & 10^{-12} & 10^{-12}\\[2ex]
{\rm g} & {\rm CAC}\,\,7 & 10^{-13} & 10^{-12} & 10^{-13}\,.
\end{array}
\]

\bigskip

From the numerical point of view, we have obtained analogous results when potentials have the form $V(x)=|x|^s$, $s>0$.

\section{Concluding remarks}

One dimensional periodic potentials are very relevant in the study of solid state systems. Yet, there are very few systems that can be exactly solved. In general, one relies in numerical and approximation methods in order to find approximate solutions to periodic systems. 

One of the methods to approximate solutions in the method of splines. However, this method requires the solution of rather complicated algebraic equations. We have shown that there is a way to circumvent this problem, which uses a variational method to fix a parameter. Then, the remainder parameters, which are coefficients of polynomials determining the spline solution, are found by a recurrence formula. This simplifies enormously the search for approximate solutions by the spline method. In addition, our method controls one of the consequences derived from the resolution of the algebraic equations: the presence of large oscillations. This good control have been shown in numerical experiments.

Thus far, we have developed a simplified method to find approximate solutions. We need to determine the eigenvalues. Here, there exists well known methods.  Starting from the Floquet theorem, we propose a determination of the eigenvalues that relies on the solution of a transcendental equation, numerically solvable. 

Finally, we have illustrated the above presentation with two examples. One is the Mathieu equation, that as remarked in the introduction, serves as a basic laboratory in order to make a first verification of the method. Since it goes well on Mathieu, we have used it to study a large class of one dimensional periodic potentials with interest in physics. In this case the potential is $V(x)=|x|^s$, $s>0$ on $[-1,1]$ and extended to the whole real line by periodicity. Finally, we have made an error analysis.

\section*{Acknowledgements}

Partial financial support is acknowledged to the Spanish MINECO (Project MTM2014-57129), the Junta de Castilla y Le on (Project VA057U16) and the Project ING 19/ i 402 of the Universidad Nacional de Rosario.

\end{document}